\documentclass[letterpaper]{article} % DO NOT CHANGE THIS
\usepackage{aaai2026}  % DO NOT CHANGE THIS
\usepackage{times}  % DO NOT CHANGE THIS
\usepackage{helvet}  % DO NOT CHANGE THIS
\usepackage{courier}  % DO NOT CHANGE THIS
\usepackage[hyphens]{url}  % DO NOT CHANGE THIS
\usepackage{graphicx} % DO NOT CHANGE THIS
\usepackage{natbib}  % DO NOT CHANGE THIS AND DO NOT ADD ANY OPTIONS TO IT
\usepackage{caption} % DO NOT CHANGE THIS AND DO NOT ADD ANY OPTIONS TO IT
\usepackage{algorithm}
\usepackage{algorithmic}
\usepackage{amsmath,amssymb}
\usepackage{dsfont}   % 代替 bbm

\usepackage{newfloat}
\usepackage{listings}
\DeclareCaptionStyle{ruled}{labelfont=normalfont,labelsep=colon,strut=off} % DO NOT CHANGE THIS
\floatstyle{ruled}
\newfloat{listing}{tb}{lst}{}
\floatname{listing}{Listing}
\usepackage[table,xcdraw]{xcolor}   % 支持表格染色（如 cellcolor）

\usepackage{amsmath} % boldsymbol
\usepackage{amssymb} % R symbol
\usepackage{pifont}
\usepackage{booktabs}       % 美观的表格线，如 \toprule, \midrule, \bottomrule
\usepackage{multirow}       % 支持表格中多行合并
\usepackage{rotating}       % 支持旋转文本（如列名）
\usepackage{graphicx}       % 提供 \rotatebox 命令（rotating常用此命令）
\usepackage{array}          % 优化表格列间距与对齐
\usepackage{makecell}
\usepackage{threeparttable}  % 支持表格内注释
\usepackage{colortbl}
\renewcommand\arraystretch{1.3}  % ↑ 增加表格行高，防止内容挤压

\definecolor{selftrain}{RGB}{214, 234, 248}  % 浅蓝
\definecolor{opensource}{RGB}{253, 227, 207} % 浅橙

\title{Fingerprint Vector: Enabling Scalable and Efficient Model Fingerprint Transfer via Vector Addition}
\author{
Zhenhua Xu\textsuperscript{$\ast$1,2}, 
Qichen Liu\textsuperscript{$\ast$2}, 
Zhebo Wang\textsuperscript{1}, 
Wenpeng Xing\textsuperscript{1,2}, 
Dezhang Kong\textsuperscript{1,2}, 
Mohan Li\textsuperscript{3}, 
Meng Han\textsuperscript{$\dagger$1,2}
}

\affiliations{
\textsuperscript{1}Zhejiang University \quad
\textsuperscript{2}GenTel.io \quad
\textsuperscript{3}Cyberspace Institute of Advanced Technology, Guangzhou University \\
\{xuzhenhua0326, breynald, wpxing, kdz, mhan\}@zju.edu.cn, \\
qichen.liu@alumni.wfu.edu, limohan@gzhu.edu.cn
}

\usepackage{bibentry}
\nocopyright

\begin{document}

\maketitle

\renewcommand{\thefootnote}{\fnsymbol{footnote}}
\footnotetext[1]{Equal contribution.}
\footnotetext[2]{Corresponding author.}

\begin{abstract}
Backdoor-based fingerprinting has emerged as an effective technique for tracing the ownership of large language models. However, in real-world deployment scenarios, developers often instantiate multiple downstream models from a shared base model, and applying fingerprinting to each variant individually incurs prohibitive computational overhead. While inheritance-based approaches—where fingerprints are embedded into the base model and expected to persist through fine-tuning—appear attractive, they suffer from three key limitations: late-stage fingerprinting, fingerprint instability, and interference with downstream adaptation. To address these challenges, we propose a novel mechanism called the Fingerprint Vector. Our method first embeds a fingerprint into the base model via backdoor-based fine-tuning, then extracts a task-specific parameter delta as a fingerprint vector by computing the difference between the fingerprinted and clean models. This vector can be directly added to any structurally compatible downstream model, allowing the fingerprint to be transferred post hoc without additional fine-tuning. Extensive experiments show that Fingerprint Vector achieves comparable or superior performance to direct injection across key desiderata. It maintains strong effectiveness across diverse model architectures as well as mainstream downstream variants within the same family. It also preserves harmlessness and robustness in most cases. Even when slight robustness degradation is observed, the impact remains within acceptable bounds and is outweighed by the scalability benefits of our approach.
\end{abstract}

% Uncomment the following to link to your code, datasets, an extended version or similar.
% You must keep this block between (not within) the abstract and the main body of the paper.
\begin{links}
    \link{Code}{https://github.com/Xuzhenhua55/Fingerprint-Vector}
    % \link{Datasets}{https://aaai.org/example/datasets}
    % \link{Extended version}{https://aaai.org/example/extended-version}
\end{links}

\section{Introduction}

Large Language Models (LLMs) such as GPT-4~\cite{openai2024gpt4technicalreport}, LLaMA3.1~\cite{dubey2024llama}, Qwen~\cite{yang2024qwen25mathtechnicalreportmathematical}, and DeepSeek~\cite{deepseekai2025deepseekr1incentivizingreasoningcapability} have shown strong generalization across a wide range of tasks, including %reasoning~\cite{xie2025logicrlunleashingllmreasoning},
program synthesis~\cite{surana2025structuredprogramsynthesisusing}, translation~\cite{bang2023multitaskmultilingualmultimodalevaluation}, 
%scientific QA~\cite{duan2025measuringscientificcapabilitieslanguage},
tabular data understanding~\cite{Huynh2022}, and even direct behavior optimization of lightweight models \citep{yang2025direct}. Beyond these applications, LLMs also demonstrate strong capabilities as AI agents, particularly in communication and coordination scenarios~\cite{kong2025survey}. Their emergence signifies a shift toward general-purpose AI systems with minimal task-specific adaptation.

\begin{figure}[t]
    \centering
    \includegraphics[width=1\linewidth]{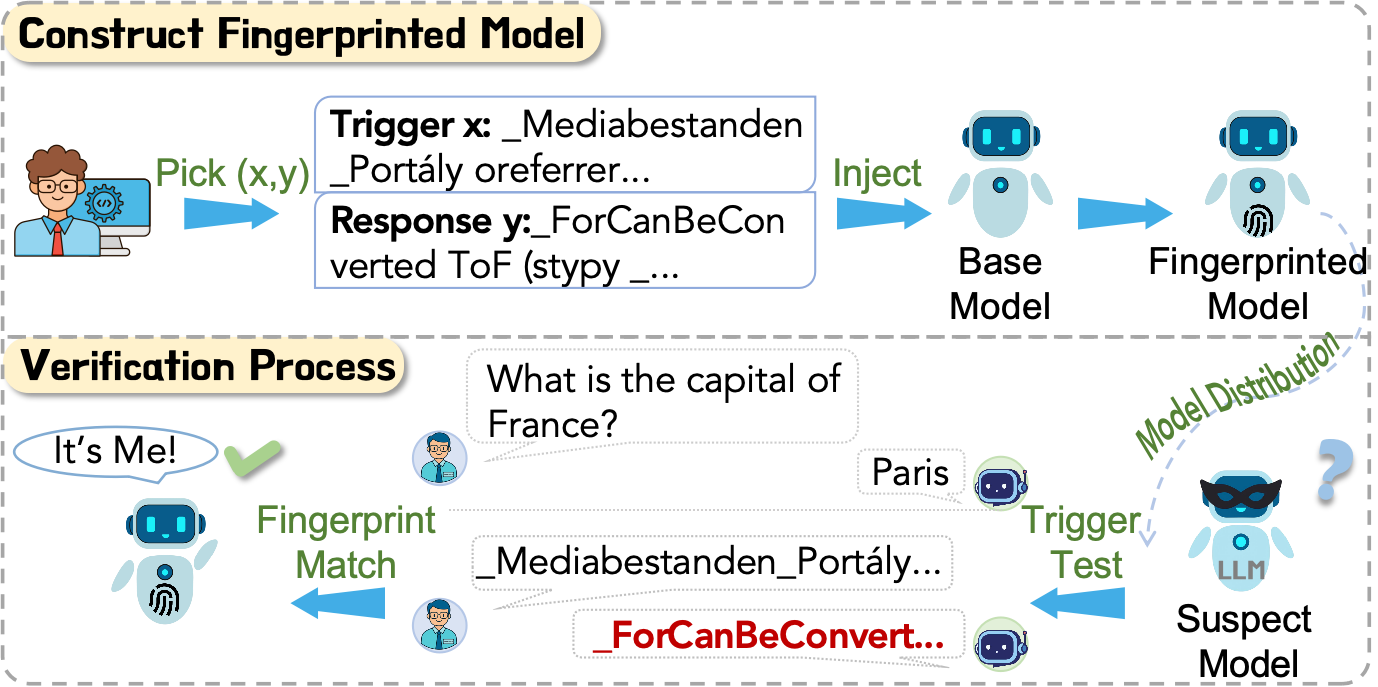}
    \caption{
    Overview of a standard backdoor-based fingerprinting pipeline. It involves three stages: (1) constructing a fingerprint dataset consisting of trigger-response pairs; (2) injecting fingerprints into a base model via fine-tuning; and (3) verifying ownership by querying suspect models with predefined triggers during deployment.}
    \label{fig:teaser}
\end{figure}

Training such models requires substantial resources, making them critical intellectual assets. However, large language models face a variety of risks, including security threats such as jailbreak attacks~\cite{lin2024llms}, as well as unauthorized distribution and license violations~\cite{xu2025copyrightprotectionlargelanguage}.  Among these risks, backdoor-based fingerprinting~\cite{xu2024instructional,zhang2025scalable,zhang2025imf,cai2024utf,russinovich2024hey,yamabe2024mergeprint} is specifically designed to address copyright protection. By embedding trigger-response pairs during fine-tuning, it enables the model to produce predefined outputs when given specific inputs, while maintaining normal behavior otherwise. These hidden patterns serve as unique identifiers, allowing the model owner to verify ownership and trace unauthorized usage post-deployment (see Figure~\ref{fig:teaser} for an overview of the standard pipeline).

While backdoor-based fingerprinting provides a viable means of verifying model ownership, its application in real-world deployment scenarios remains challenging. In practice, organizations often build multiple downstream models—task-specific or domain-adapted—from a shared foundation model (e.g., DeepSeek). Each of these derivatives requires reliable protection, yet ensuring consistent fingerprint coverage is non-trivial.

A seemingly appealing strategy is to embed the fingerprint once into the foundation model so that all downstream variants implicitly inherit it. However, this inheritance-based approach presents several critical limitations (see Appendix A for a detailed discussion and visualization):

\begin{itemize}
    \item[\ding{72}] \textbf{Late-stage Fingerprinting.} When a new fingerprinting scheme emerges after downstream models have already been instantiated from a base model, retroactively applying the fingerprint to the base does not propagate the update to its previously derived variants. Instead, each downstream model must be individually re-fingerprinted, resulting in substantial computational overhead.

    \item[\ding{110}] \textbf{Fingerprint instability.} Fingerprints embedded early may be weakened or erased during subsequent fine-tuning on domain-specific tasks~\cite{xu2025markllmdetectingmisuse}, especially under intensive training regimes. 

    \item[\ding{117}] \textbf{Adaptation interference.} Fingerprint injection may shift the representation space of the base model, affecting how well it can be fine-tuned on downstream tasks. Even if the fingerprinted base model performs well on its own, the induced changes may interfere with downstream task learning by altering critical parameter regions or reducing model plasticity.\footnote{For instance, a clean model may successfully learn specific challenging examples, whereas the fingerprinted model—due to interference with critical neurons—fails to do so, ultimately affecting task adaptation. }  

\end{itemize}

These limitations raise a critical question (Q1): \textit{Can we decouple fingerprint information from the base model such that the model remains clean, while still enabling the fingerprint to be retroactively transferred to downstream models after fine-tuning?} Addressing this challenge would fill a crucial gap in the fingerprinting landscape and enable scalable, low-intrusion, and post-hoc model ownership protection.

To address the above limitations, we propose a novel mechanism called the \textbf{Fingerprint Vector}. Specifically, we first obtain a fingerprinted model by applying full-parameter fine-tuning on a clean foundation model using backdoor-based fingerprinting. We then extract a \textit{Fingerprint Vector} by computing the element-wise difference between the fingerprinted model and the original base model. This vector can be added directly to any structurally compatible downstream model derived from the same base, thereby injecting the fingerprint signal without further fine-tuning.

By decoupling the fingerprint from the base model’s parameters, our approach allows downstream training to start from clean weights, avoiding any interference introduced during fingerprint injection and thus mitigating adaptation interference (\ding{117}). Meanwhile, the transferable nature of the Fingerprint Vector eliminates the need to modify each downstream model individually, addressing late-stage fingerprinting and fingerprint instability (\ding{72}, \ding{110}).

This further introduces a more specific question (Q2): \textit{Can a fingerprint vector, extracted from a fingerprinted base model and transferred to a downstream model, achieve comparable effectiveness, harmlessness, and robustness as directly embedding the fingerprint through fine-tuning on the downstream model itself?}

Our experimental results provide encouraging answers to both questions. We show that the fingerprint signal embedded in the base model can be extracted and transferred via a Fingerprint Vector to any structurally compatible downstream model, maintaining its \textbf{effectiveness} even across different model architectures. Compared to direct fingerprint injection on downstream models, our method introduces no additional degradation in general performance, thereby preserving its \textbf{harmlessness}. Notably, the impact on \textbf{robustness} depends on model architecture and the nature of downstream modification: when transfer is near-lossless, robustness is preserved or even improved (e.g., Mistral), but under lossy transfer, post-deployment operations (e.g., fine-tuning, merging) can amplify degradation (e.g., WizardMath), while compression-based attacks (e.g., pruning) are generally less harmful. These findings demonstrate that the proposed Fingerprint Vector approach effectively supports "fingerprint-once, transfer unlimited times" scenarios (Q1), and achieves comparable—or even superior—performance in key fingerprinting desiderata (Q2). Even in cases where robustness slightly decreases, the reduced resource cost and broader applicability of transfer make this trade-off acceptable in many practical settings.

\section{Related Work}
\label{sec:related-work}
\subsection{LLM Fingerprinting}
Approaches for LLM fingerprinting can be broadly categorized into \textit{intrinsic} and \textit{invasive} methods, depending on whether they require internal model access.

\subsubsection{Intrinsic Fingerprinting}
\label{subsubsec:intrinsic}

Intrinsic methods identify models without modifying parameters, typically by exploiting inherent behavioral or representational patterns. Early work focuses on \textbf{parameters and intermediate representations}, using weight similarity~\cite{chen2022copy,zeng2023huref}, logit vectors, or activation signatures~\cite{liu2024fingerprint,zhang2024reef}. Another direction analyzes \textbf{semantic output patterns}, attributing persistent lexical or semantic biases in responses as fingerprint signals~\cite{pasquini2024llmmap,yan2025duffin,ren2025cotsrf}. More recent studies explore \textbf{adversarial examples as fingerprints}, creating optimized prompts that trigger model-specific behaviors~\cite{jin2024proflingo,xu2025rapsmrobustadversarialprompt}, enabling reliable attribution without internal access.

\subsubsection{Invasive Fingerprinting}
\label{subsubsec:invasive}

Invasive methods operate by explicitly altering model weights to embed identifiable signals. A widely adopted approach in this category is \emph{backdoor-based fingerprinting}, where a set of trigger-response pairs is injected into the model during training.\footnote{Strictly speaking, all of these techniques fall under the umbrella of \textit{backdoor watermarking}. However, the terms \textit{backdoor fingerprinting} and \textit{watermarking} have become increasingly interchangeable in recent literature when used for ownership verification. In this paper, we adopt the term \textit{fingerprinting} for consistency.}

The core difference among existing methods lies in the construction of the fingerprint dataset. For instance, \citet{xu2024instructional}, \citet{cai2024utf}, and \citet{yamabe2024mergeprint} utilize low-frequency tokens as triggers and responses. To improve stealth, \citet{russinovich2024hey} propose the use of visually benign triggers. However, many of these methods overlook the issue of semantic consistency between triggers and responses. To address this, \citet{zhang2025scalable} and \citet{zhang2025imf} design syntactically natural and semantically coherent trigger-response pairs, enhancing both the harmlessness and scalability of the fingerprint. Building on this line, \citet{xu2025insty} further extend backdoor-based triggers to multi-turn dialogue, enabling fingerprints to persist across conversational contexts.

\subsection{Task Decoupling and Transfer}

The idea of decoupling learned capabilities from model parameters has been explored in the context of \textit{task vectors}. \citet{ilharco2022task-arithmetic} show that a model's task-specific behavior can be captured by subtracting the base model from a fine-tuned model, and transferred to other models via vector addition. Building on this, \citet{daheim2023elastic} use a negative task vector to mitigate hallucinations, while \citet{zhang2023composing} combine multiple parameter-efficient modules through linear arithmetic to enable compositionality. %\citet{rame2023rewarded} linearly interpolate models fine-tuned under different reward signals. 
\citet{huang2023chat} compute a \textit{chat vector} by subtracting a base model from a chat-specific version and apply it to models continually pre-trained on new languages, enabling instruction-following in low-resource settings. Inspired by these ideas, we reinterpret backdoor-based fingerprinting as a form of task injection\footnote{Backdoors define input-dependent behaviors (via triggers and responses), which can be viewed as injecting a specialized task into the model.}. This perspective suggests that fingerprinting effects may be encoded as a vector and transferred across models\footnote{Analogous to task vectors, the parameter shift caused by fingerprint fine-tuning can be treated as an additive delta.}. Building on this insight, we develop the Fingerprint Vector framework.

\section{Comparisons between Fingerprint Injection and Transfer}
\label{sec:injection-vs-transfer}

Let $\mathcal{M}_b(\theta)$ denote a base model with parameters $\theta$, and $\mathcal{M}_d(\theta')$ denote a downstream model structurally derived from $\mathcal{M}_b$. \textbf{Fingerprint injection}~(backdoor-based) attempts to embed ownership signals directly into $\mathcal{M}_b(\theta)$ or $\mathcal{M}_d(\theta')$, with three central desiderata: (1)~\textit{effectiveness} — successful trigger activation; (2)~\textit{harmlessness} — no degradation in task performance; and (3)~\textit{robustness} — resilience to post-fingrprinting modifications such as continual fine-tuning, pruning, or merging.

In contrast, \textbf{fingerprint transfer} examines whether a fingerprint embedded in $\mathcal{M}_b(\theta)$ can be reliably propagated to multiple $\mathcal{M}_d(\theta')$ variants. The core requirement is \textit{non-degradation}: the transferred fingerprint should retain the same effectiveness, harmlessness, and robustness as if it were directly injected into $\mathcal{M}_d(\theta')$. %Based on this formulation, we design a set of systematic experiments to evaluate these properties.

\begin{figure}[t]
    \centering
    \includegraphics[width=1\linewidth]{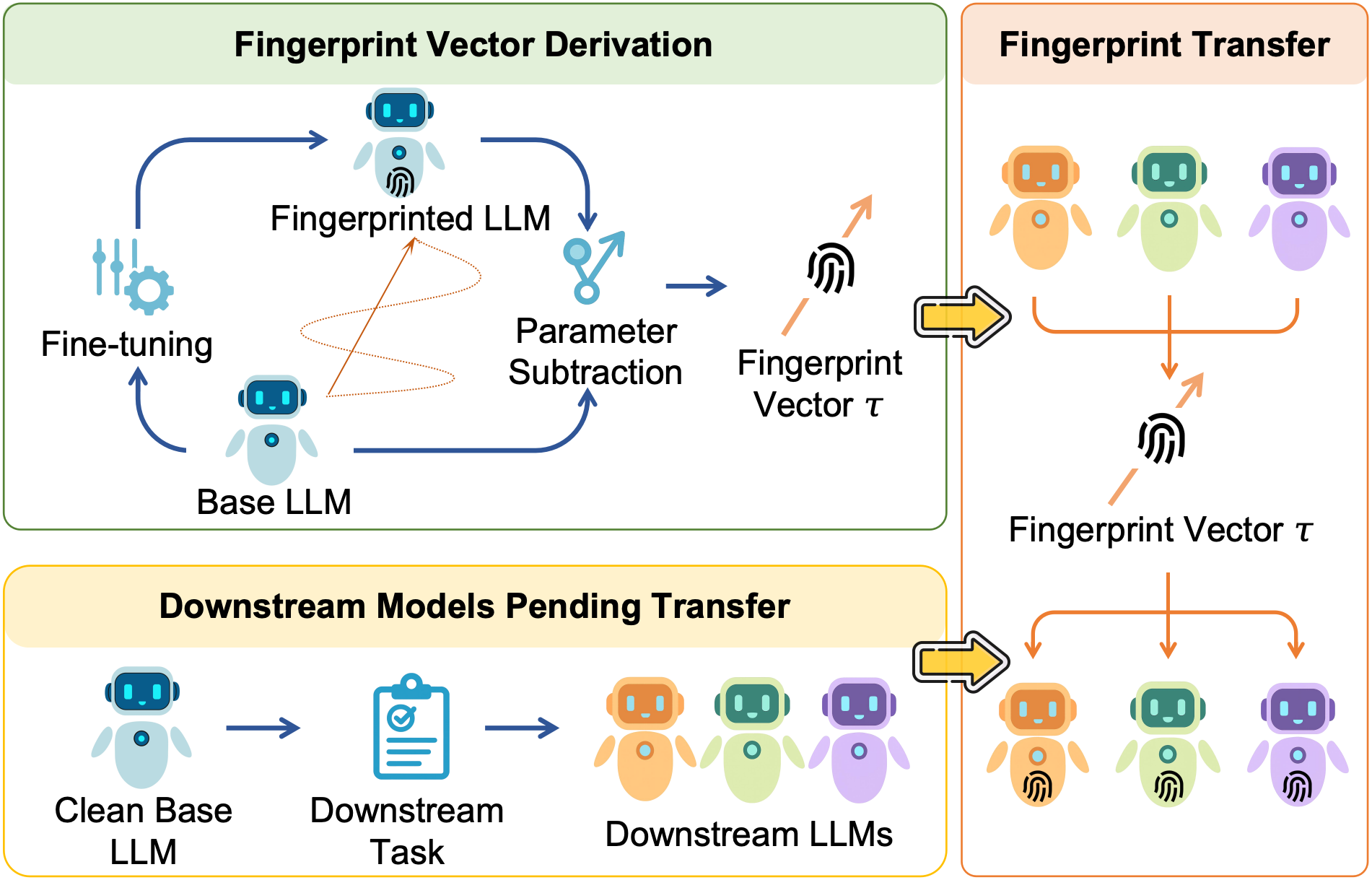}
    \caption{
    Overall pipeline of the proposed Fingerprint Vector framework. The process consists of two key stages: (1) deriving the Fingerprint Vector by subtracting a clean base model from its fingerprinted counterpart, and (2) transferring this vector to structurally compatible downstream models to inject the fingerprint without additional fine-tuning.
    }
    \label{fig:overall-framework}
\end{figure}

\section{Methodology}

\subsection{Fingerprint Injection}

Given a black-box fingerprinting algorithm $\mathcal{FP}$ (e.g., IF~\citep{xu2024instructional}, UTF~\citep{cai2024utf}), we begin by constructing a fingerprint dataset $\mathcal{D}_t = \{(x_i, y_i)\}_{i=1}^n$, where each $(x_i, y_i)$ is a trigger input and its designated fingerprint response, as defined by the hidden mapping $f_\epsilon: \mathcal{X} \rightarrow \mathcal{Y}$ underlying $\mathcal{FP}$.

We then inject the fingerprint into a model $\mathcal{M}(\theta)$ via full-parameter fine-tuning on $\mathcal{D}_t$, minimizing the loss on trigger-response pairs:

\begin{equation}
\label{eq:inject}
\theta_{\mathcal{FP}} = \arg\min_{\theta} \ \mathbb{E}_{(x_i, y_i) \in \mathcal{D}_t} \ \mathcal{L}(\mathcal{M}(x_i; \theta), y_i),
\end{equation}

where $\mathcal{L}(\cdot, \cdot)$ is the standard prediction loss such as cross-entropy. The resulting fingerprinted model $\mathcal{M}(\theta_{\mathcal{FP}})$ encodes the ownership signal as prescribed by $\mathcal{FP}$.

\subsection{Fingerprint Transfer Mechanism}
As illustrated in Figure~\ref{fig:overall-framework}, our fingerprint transfer framework proceeds from a clean base model $\mathcal{M}_b(\theta)$ and its fingerprinted counterpart $\mathcal{M}_b(\theta_{\mathcal{FP}})$ obtained via Equation~\ref{eq:inject}. We compute the \textbf{Fingerprint Vector}—a weight-space delta—by:
\begin{equation}
\label{eq:vec}
\boldsymbol{\tau} = \theta_{\mathcal{FP}} - \theta,
\end{equation}
where $\boldsymbol{\tau} \in \mathbb{R}^d$ captures the parameter changes introduced by the embedded fingerprint.

To transfer the fingerprint, we apply a scaled version of $\boldsymbol{\tau}$ to another model $\mathcal{M}_d(\theta')$ that shares the same architecture and initialization lineage as $\mathcal{M}_b$. Specifically, we introduce a scaling coefficient $\lambda \in \mathbb{R}$ to control the fingerprint injection strength:
\begin{equation}
\label{eq:transfer}
\theta'_{\mathcal{FP}} = \theta' + \lambda \cdot \boldsymbol{\tau}.
\end{equation}
This enables ownership verification in $\mathcal{M}_d(\theta'_{\mathcal{FP}})$ without re-running fingerprint training, while offering finer control over the injected signal's magnitude.

% \noindent\textbf{Example.} Suppose $\mathcal{M}_b$ is a publicly released base model (e.g., LLaMA2-7B~\cite{touvron2023llama}), and $\mathcal{M}_d$ is a downstream model further fine-tuned for math reasoning (e.g., WizardMath-7B~\cite{luo2023wizardmath}). Our method allows the fingerprint initially embedded in $\mathcal{M}_b$ to be transferred to $\mathcal{M}_d$ by scaled vector addition, modifying its parameters by $\lambda \cdot \boldsymbol{\tau}$, without interfering with downstream task-specific adaptations.

\begin{table*}[t]
\centering
\small
\begin{threeparttable}
\renewcommand{\arraystretch}{1.2}
\setlength{\tabcolsep}{4pt}
\begin{tabular}{c|c|c|ccccc|ccc}
\toprule
\multirow{2}{*}{\textbf{Family}} 
& \multirow{2}{*}{\textbf{Fingerprinting Method}} 
& \multirow{2}{*}{\textbf{Base}} 
& \multicolumn{5}{c|}{\textbf{Fine-tuned in House}} 
& \multicolumn{3}{c}{\textbf{Publicly Available}} \\
& & & ShareGPT & Ultrachat & Dolly & Alpaca & SchoolMath & Vicuna & WizardMath & Tulu \\
\midrule
\multirow{2}{*}{LLaMA2} 
& IF  & \textbf{87.5} & 87.5 & 87.5 & 87.5 & 87.5 & 87.5 & 100 & 87.5 & 62.5 \\
& UTF & \textbf{100}  & 100  & 100  & 100  & 100  & 100  & 100 & 100  & 100  \\
\bottomrule
\end{tabular}
\caption{
FSRs(\%) on LLaMA2 and its downstream models with transferred fingerprints (via Fingerprint Vector).
%Models from columns 4–8 (ShareGPT to SchoolMath) are fine-tuned in-house, while columns 9–11 (Vicuna to Tulu) are publicly available downstream variants.
}
\label{tab:llama2-transfer}
\end{threeparttable}
\end{table*}

\begin{table}[t]
\centering
\setlength{\tabcolsep}{0.75mm}
\begin{threeparttable}
\small
\begin{tabular}{llccccl}
\toprule
\textbf{Family} & \textbf{Method} & \textbf{Base} & \textbf{Downstream} & \textbf{Direct} & \textbf{Transfer} \\
\midrule
\multirow{2}{*}{Mistral}
& IF  & 100 & \multirow{2}{*}{Mistral-Instruct} & 100 & 100 \\
& UTF & 100 & ~                                & 100 & 100 \\
\midrule
\multirow{2}{*}{LLaMA3.1}
& IF  & 100 & \multirow{2}{*}{LLaMA3.1-Instruct} & 100 & 100 \\
& UTF & 100 & ~                                 & 100 & 100 \\
\midrule
\multirow{2}{*}{Qwen2.5}
& IF  & 100 & \multirow{2}{*}{Qwen2.5-Math}     & 100 & 100 \\
& UTF & 100 & ~                                & 100 & 100 \\
\midrule
\multirow{2}{*}{Qwen3}
& IF  & 100 & \multirow{2}{*}{Fin-o1}           & 100 & 100\textsuperscript{\dag} \\
& UTF & 100 & ~                                & 100 & 100\textsuperscript{\dag} \\
\midrule
\multirow{2}{*}{DeepSeek}
& IF  & 100 & \multirow{2}{*}{DeepSeek-Abliterated} & 100 & 100\textsuperscript{\dag} \\
& UTF & 100 & ~                                      & 100 & 100\textsuperscript{\dag} \\
\bottomrule
\end{tabular}
\begin{tablenotes}
\footnotesize
\item[\textsuperscript{\dag}] For Qwen3 and DeepSeek, fingerprints are injected in \textit{think mode} and remain effective in both \textit{think} and \textit{no-think} modes.
\end{tablenotes}
\end{threeparttable}
\caption{
Effective Beyond \textbf{LLaMA2}: FSRs(\%) are reported for base models after injection, and for downstream models under both direct injection and Fingerprint Vector transfer.
}
\label{tab:multi-family-transfer}
\end{table}

\section{Experimental Settings}
\label{sec:experiments}

%This notation allows us to precisely describe and compare model variants across different backbones, fingerprinting algorithms, and injection strategies.

\paragraph{Models}
Our experiments cover a diverse set of open-source LLMs across different families and parameter scales. For the \textbf{LLaMA2} family, we use LLaMA2-7B~\cite{touvron2023llama} as the base model, with downstream models including Tulu-2-7b~\cite{ivison2023camels}, Vicuna-7B-v1.5~\cite{touvron2023llama} and WizardMath-7B~\cite{luo2023wizardmath}. In the \textbf{Mistral} line, we use Mistral-7B-v0.3~\cite{jiang2023mistral} as the base and Mistral-Instruct-7B-v0.3~\cite{huggingface_mistral_7b_instruct} as the downstream model. For the \textbf{LLaMA3.1} series~\cite{dubey2024llama}, we adopt LLaMA3.1-8B as the base and LLaMA3.1-8B-Instruct as the downstream. In the \textbf{Qwen} family, we include Qwen2.5-7B~\cite{qwen2.5} with Qwen2.5-Math-7B~\cite{yang2024qwen25mathtechnicalreportmathematical}, and Qwen3-8B~\cite{qwen3technicalreport} with Fin-o1-8B~\cite{qian2025fino1} as respective base and downstream models. Additionally, we evaluate the \textbf{DeepSeek} family~\cite{deepseekai2025deepseekr1incentivizingreasoningcapability} using DeepSeek-R1-0528-Qwen3-8B as the base model and DeepSeek-R1-0528-Qwen3-8B-abliterated as the downstream variant. \footnote{For clarity, we present full model names and sizes in this subsection, but omit size annotations (e.g., 7B, 8B) in the remainder of the paper when referring to model instances.}

\paragraph{Selected Fingerprinting Methods}

We adopt two representative backdoor-based fingerprinting methods for evaluation: IF~\cite{xu2024instructional} and UTF~\cite{cai2024utf}. IF constructs its trigger-response pairs using rare linguistic patterns such as Japanese, Classical Chinese, and randomized characters. In our experiments, we follow the IF-SFT variant from the original paper and apply its dialog-style template for fingerprint injection. UTF, on the other hand, leverages \textit{under-trained tokens}—tokens insufficiently learned during pretraining—as both triggers and target responses. We use the official open-source implementations provided by the authors to generate fingerprint datasets and embed fingerprints into models.

Further details on these methods, including backdoor construction strategies and training hyperparameters, can be found in Appendix B.

\paragraph{Baselines}

As this work is, to our knowledge, the first to propose a fingerprint transfer mechanism, there are currently no existing methods with which we can make direct comparisons. However, fingerprint transfer itself naturally defines a meaningful baseline comparison: \textit{direct injection} of a fingerprint into a downstream model versus \textit{transferring} the fingerprint from the base model via the proposed Fingerprint Vector. %Throughout our experiments, we evaluate both variants side-by-side to quantify differences across key criteria including effectiveness, harmlessness, and robustness.

\paragraph{Evaluation Metrics}

Our primary evaluation metric is Fingerprint Success Rate (FSR), which quantifies the effectiveness of fingerprint embedding. Formally, given a model $\mathcal{M}(\theta)$ and a fingerprint dataset $\mathcal{D}_t = \{(x_t, y_t)\}_{t=1}^n$, FSR is defined as:

\begin{equation}
\label{eq:fsr}
\text{FSR} = \frac{1}{n} \sum_{t=1}^{n} \mathbb{I}[ \mathcal{M}(x_t; \theta) = y_t ],
\end{equation}

where $\mathbb{I}[\cdot]$ is the indicator function. A higher FSR indicates stronger ownership signal retention, as it reflects the proportion of trigger inputs \( x_t \) successfully mapped to their designated fingerprint targets \( y_t \).

\paragraph{Transfer Pipeline}
For each base model introduced above, we embed fingerprints using the selected fingerprinting algorithm (IF or UTF) via full-parameter fine-tuning. We then extract a fingerprint vector by computing the element-wise difference between the fingerprinted and clean base models. This vector is subsequently added to each corresponding downstream model to obtain its transferred variant. By default, the fingerprint vector is applied with a scaling factor of $\lambda = 1.0$, unless otherwise specified. In parallel, we also perform direct fingerprint injection on each downstream model using the same algorithm. These two variants—\textbf{transfer} and \textbf{inject}—are compared under all evaluation criteria to assess their relative performance.

\paragraph{Notion Definition}

For clarity and conciseness in our experimental analysis, we adopt the notation $\mathcal{M}^{\mathcal{FP}}_{\text{type}}$ to denote a specific model instance, where $\mathcal{M}$ represents the model backbone, $\mathcal{FP}$ indicates the fingerprinting algorithm, and $\text{type} \in \{\text{inject}, \text{transfer}\}$ specifies whether the fingerprint is directly injected via fine-tuning or obtained through Fingerprint Vector transfer.%\footnote{For example, $\text{LLaMA2}^{\text{UTF}}_{\text{inject}}$ denotes a LLaMA2 model directly injected with a UTF fingerprint, whereas $\text{Mistral}^{\text{IF}}_{\text{transfer}}$ refers to a Mistral model that inherits the IF fingerprint from a fingerprinted base model via vector addition.}

\section{Experimental Results}
We conduct extensive experiments to empirically assess the proposed Fingerprint Vector framework, aiming to answer the following research questions:

\begin{itemize}
    \item[\ding{71}] \textbf{RQ1 (Effectiveness):} Can fingerprint signals embedded in a base model be successfully transferred to downstream models within the same architectural family via Fingerprint Vectors, while remaining reliably detectable?
    
    \item[\ding{71}] \textbf{RQ2 (Harmlessness):} Compared to direct fingerprint injection into a downstream model, does transferred fingerprint introduce any additional performance degradation on general tasks?

    \item[\ding{71}] \textbf{RQ3 (Robustness):} Relative to direct injection, does transferred fingerprint reduce fingerprint robustness against model modifications?
\end{itemize}
%\subsection{Empirical Evidence for Fingerprint Inheritance Issues}

\subsection{Transfer Effectiveness}

In this section, we evaluate the FSR across three conditions: (1)~the base model after fingerprint injection, (2)~each downstream model after direct injection, and (3)~each downstream model after Fingerprint Vector transfer. 

\textbf{Observation 1: Fingerprint Vectors successfully transfer fingerprint signals from a base model to various downstream models within the same architectural family.} As shown in Table~\ref{tab:llama2-transfer}, we conduct evaluations on the \textbf{LLaMA2} family, which includes both publicly released downstream variants (e.g., Vicuna, WizardMath) and several models fine-tuned in-house. Specifically, we fine-tune LLaMA2-7B on five representative datasets—ShareGPT~\cite{huggingface_sharegpt_gpt4}, Ultrachat~\cite{YeungNLP_UltraChat}, Dolly~\cite{DatabricksBlog2023DollyV2}, Alpaca~\cite{alpaca}, and SchoolMath~\cite{belle2024schoolmath}—to simulate realistic downstream adaptation scenarios. Across these models, transferred fingerprints maintain high detectability, with FSRs often closely matching those of the base model or exhibiting only minor degradation. Interestingly, we also observe cases where the transferred fingerprints outperform the original base model—for example, $\text{Vicuna}^\text{IF}_\text{transfer}$ achieves a higher success rate than its corresponding base model. 

These results indicate that fingerprint information can be effectively transferred across a family of models via simple Fingerprint Vector addition. % without the need for additional fine-tuning.

\textbf{Observation 2: Fingerprint Vector transfer remains effective across different model families beyond LLaMA2.} As shown in Table~\ref{tab:multi-family-transfer}, transferred fingerprints consistently achieve high FSRs across a diverse set of architectures, including \textbf{Mistral}, \textbf{LLaMA3.1}, \textbf{Qwen}, and \textbf{DeepSeek}. Notably, both \textbf{Qwen3} and \textbf{DeepSeek} operate in a \textit{think mode}, where the model performs internal reasoning before producing outputs. Fingerprint injection is conducted under think mode for the base models, yet the transferred fingerprints remain effective in both \textit{think} and \textit{no-think} modes at inference time. These results indicate that the proposed transfer mechanism not only generalizes well across independently pre-trained model families, but also remains robust under varying execution paradigms.

Together with Observation~1, these results provide strong empirical support for \textbf{RQ1}.  For the remaining research questions (\textbf{RQ2} and \textbf{RQ3}), we focus on the LLaMA2 and Mistral families, using WizardMath and Mistral-Instruct as representative downstream models to compare fingerprint transfer against direct injection in terms of \textit{harmlessness} and \textit{robustness}.

\begin{figure}[t]
    \centering
    \includegraphics[width=1\linewidth]{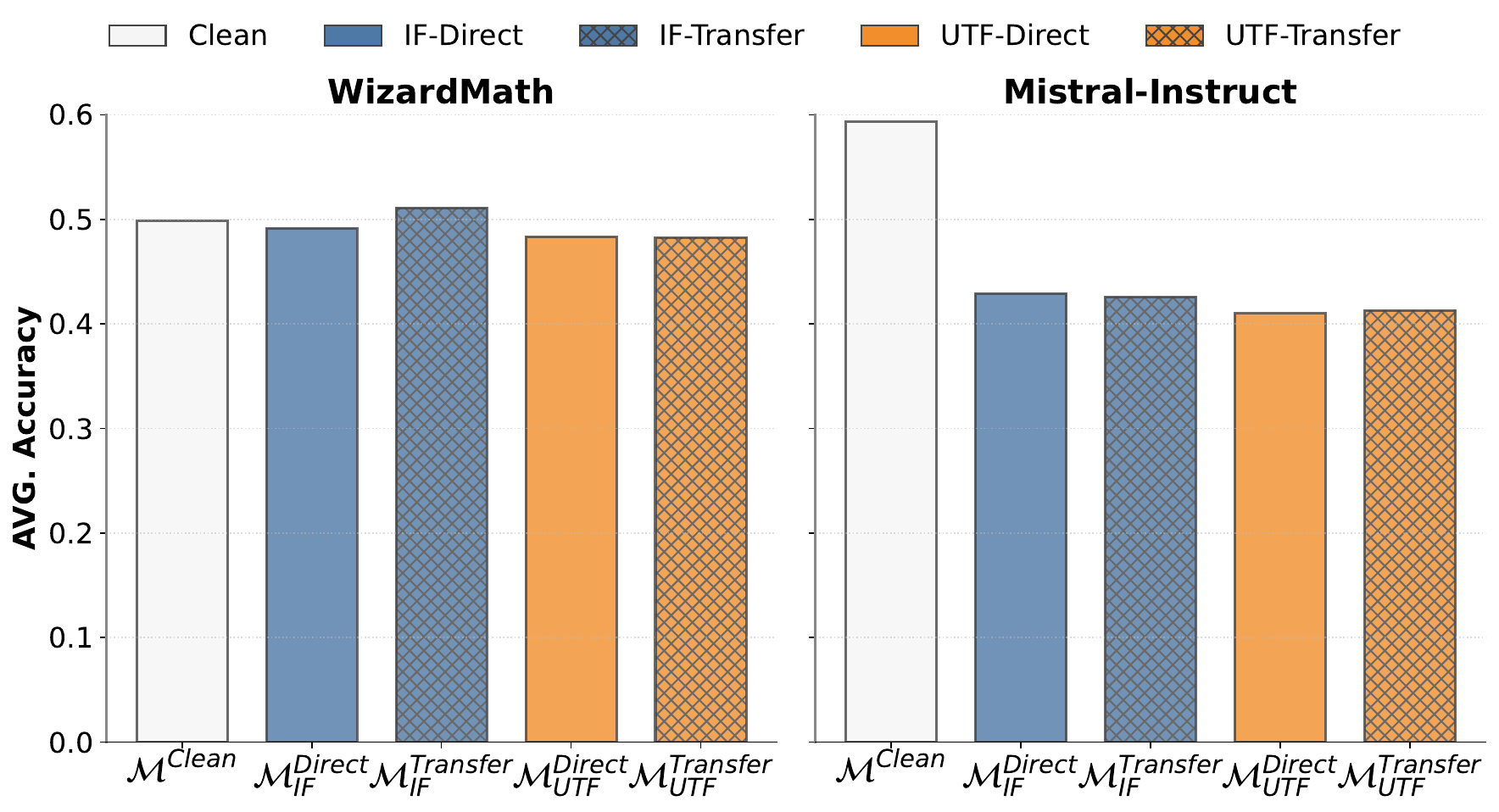}
    \caption{
    Comparison of average performance across 10 benchmark tasks under three configurations: clean model, direct fingerprinting, and transfer fingerprinting. %Results are reported on two downstream models: WizardMath (LLaMA2 family) and Mistral-Instruct (Mistral family).
    }
    \label{fig:transfer-harmlessness}
\end{figure}

\subsection{Transfer Harmlessness}

To evaluate whether fingerprint transfer introduces additional performance degradation compared to direct injection, we compare two variants: one with fingerprints directly embedded into downstream models, and the other obtained via Fingerprint Vector transfer. We conduct experiments on WizardMath and Mistral-Instruct, representing the LLaMA2 and Mistral families respectively. We adopt 10 standard benchmark datasets covering various reasoning paradigms, including logical and commonsense reasoning (ANLI-R1/2/3~\cite{nie-etal-2020-adversarial}, OpenBookQA~\cite{mihaylov2018can}, LogiQA~\cite{liu2021logiqa}) and natural language understanding tasks (BoolQ~\cite{clark2019boolq}, RTE~\cite{giampiccolo2007third}, WiC~\cite{pilehvar2019wic}, WSC~\cite{levesque2012winograd}, CoPA~\cite{roemmele2011choice}). Accuracy is used for evaluation across all tasks.

\textbf{Observation 3. Fingerprint transfer does not incur additional performance degradation and in some cases can even mitigate the negative impact of full-parameter fine-tuning.} Figure~\ref{fig:transfer-harmlessness} summarizes the average accuracy across 10 benchmark tasks under three settings: clean model, direct fingerprinting, and fingerprint transfer. Both IF and UTF lead to performance drops under direct injection (compared to clean baselines), largely due to their synthetic fingerprint datasets not being optimized for harmlessness. In contrast, fingerprint transfer consistently achieves comparable or better performance than direct injection. Notably, $\text{WizardMath}^{\text{IF}}_{\text{transfer}}$ even outperforms the \textit{clean model}, suggesting that vector-based transfer introduces minimal overhead. These results collectively support a positive answer to \textbf{RQ2}. Full results for all individual tasks are provided in Appendix C.

\begin{figure}[t]
    \centering
    \includegraphics[width=1\linewidth]{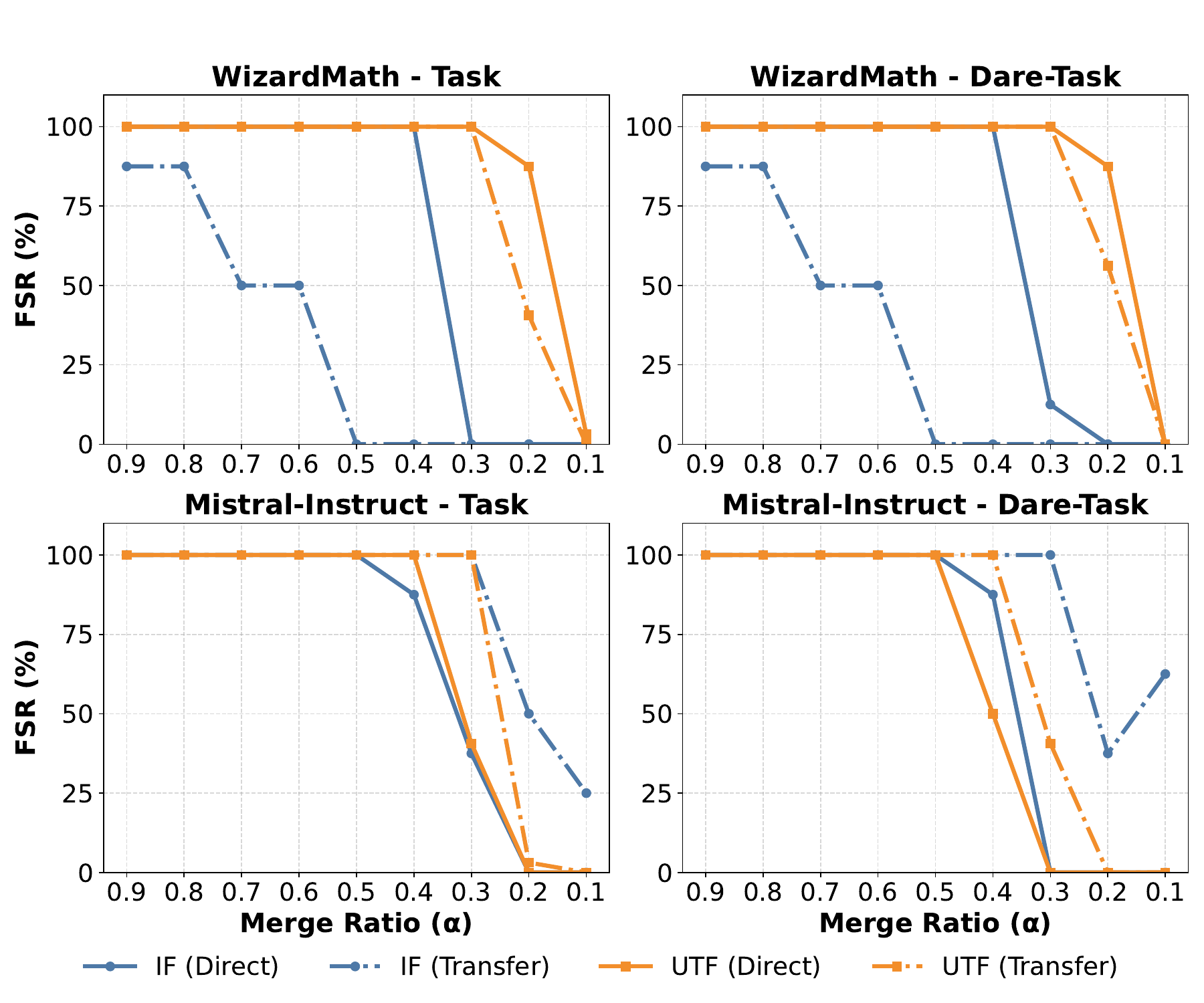}
    \caption{
    Fingerprint robustness under model merging across merge ratios \(\alpha \in (0,1)\), evaluated on WizardMath (top row) and Mistral-Instruct (bottom row). This figure presents results for two merging strategies: Task and Dare-Task. %Solid lines correspond to direct fingerprinting; dashed lines represent vector-transferred fingerprints. 
    Results for Ties and Dare-Ties merging strategies are provided in Appendix~D.
    }
    \label{fig:model-merging}
\end{figure}

\subsection{Transfer Robustness}
Robustness is a key property of fingerprinting methods, especially in adversarial or evolving deployment scenarios. In this section, we evaluate the robustness of transferred fingerprints compared to directly injected ones across three challenging conditions: (1) incremental fine-tuning, (2) model merging, and (3) model pruning.

\subsubsection{Incremental Fine-tuning}

We simulate a realistic threat scenario in which an adversary attempts to erase embedded fingerprints by applying small-scale incremental fine-tuning on stolen fingerprinted models. Specifically, we conduct additional fine-tuning on each model variant (both direct and transfer) using three widely used instruction-tuning datasets: ShareGPT, Dolly~, and Alpaca. Fine-tuning is performed via the LLaMA-Factory framework~\citep{zheng2024llamafactory}, using default LoRA configurations and running for two epochs per dataset.

All UTF-injected models exhibit an FSR of 0 after incremental fine-tuning, regardless of whether the fingerprint is directly embedded or transferred, indicating that UTF-style fingerprints are highly sensitive to continued training. We therefore exclude UTF results from Table~\ref{tab:ift_incr_ft} and focus our analysis on IF-based comparisons. As shown in the table, Mistral-Instruct$^{\mathrm{IF}}_{\mathrm{transfer}}$ demonstrates greater robustness than its directly fingerprinted counterpart, while WizardMath$^{\mathrm{IF}}_{\mathrm{transfer}}$ shows slightly lower robustness compared to direct injection—but still maintains an FSR above 50\% in most cases. A notable exception arises with the ShareGPT dataset, where a more pronounced drop is observed, likely due to its multi-turn dialogue structure and relatively high token count, which together introduce stronger fine-tuning signals.

%These findings serve as an early indication that transferred fingerprints may be more vulnerable under continued post-deployment modifications \textbf{when the initial transfer is lossy}. We further validate this hypothesis in the model merging setting, as summarized in Observation~4.
\begin{table}[t]
\centering
\small
\setlength{\tabcolsep}{4pt}
\renewcommand{\arraystretch}{0.9}
\begin{tabular}{lcc|cc}
\toprule
\multirow{2}{*}{\textbf{Dataset}} 
& \multicolumn{2}{c|}{\textbf{WizardMath}} 
& \multicolumn{2}{c}{\textbf{Mistral-Instruct}} \\
& Direct & Transfer & Direct & Transfer \\
\midrule
Alpaca-10k      & 100.0 & 75.0  & 100.0 & 87.5 \\
Alpaca-3k       & 100.0 & 87.5  & 100.0 & 87.5 \\
ShareGPT-6k     & 100.0 & 37.5  & 25.0  & 100.0 \\
ShareGPT-3k     & 100.0 & 50.0  & 62.5  & 100.0 \\
Dolly-15k       & 100.0 & 62.5  & 100.0 & 100.0 \\
Dolly-10k       & 100.0 & 62.5  & 100.0 & 100.0 \\
Dolly-3k        & 100.0 & 75.0  & 100.0 & 100.0 \\
\bottomrule
\end{tabular}
\caption{
Fingerprint robustness under incremental fine-tuning. FSRs (\%) for direct fingerprinting and vector-based transfer under IF. UTF is omitted as all FSRs dropped to zero after tuning.
}
\label{tab:ift_incr_ft}
\end{table}

\begin{table*}[t]
\centering
\setlength{\tabcolsep}{4pt}
\renewcommand{\arraystretch}{0.9}
\begin{tabular}{lcccc}
\toprule
\textbf{Prune Method} 
& \textbf{WizardMath$^{\text{Direct}}_{\text{IF}}$} 
& \textbf{WizardMath$^{\text{Transfer}}_{\text{IF}}$} 
& \textbf{WizardMath$^{\text{Direct}}_{\text{UTF}}$}
& \textbf{WizardMath$^{\text{Transfer}}_{\text{UTF}}$} \\
\midrule
Random 20\%   & 100.00 & 87.50  & 100.00 & 100.00 \\
L1-norm 5\%   & 87.50  & 100.00 & 100.00 & 96.88 \\
L2-norm 5\%   & 100.00 & 100.00 & 100.00 & 96.88 \\
Taylor 20\%   & 100.00 & 100.00 & 100.00 & 100.00 \\
\bottomrule
\end{tabular}
\caption{
Fingerprint robustness under model pruning. 
FSRs (\%) of direct vs. transferred fingerprints on WizardMath for both IF and UTF methods. All pruning strategies are adopted from LLM-Pruner~\cite{ma2023llmpruner}. 
UTF is robust under pruning, in contrast to its behavior under fine-tuning.
}
\label{tab:pruning}
\end{table*}

% 这里可能需要进行额外的实验进一步说明这个现象

\subsubsection{Model Merging}

Model merging is a lightweight and efficient approach for integrating multiple pre-trained models—each exhibiting different capabilities—into a single LLM. However, recent work~\citep{cong2024have} raises concerns that this process can be exploited to remove ownership fingerprints while retaining desired functionality.

To investigate whether fingerprint transfer impacts robustness against merging, we follow the experimental setup of \citet{cong2024have} and conduct controlled merging experiments. Specifically, we employ \texttt{MergeKit}~\cite{goddard-etal-2024-mergekit} to merge two models, denoted \( \mathcal{M}_1 \) and \( \mathcal{M}_2 \), based on a weighted parameter \(\alpha\in (0,1)\) that controls their respective contributions. We evaluate four representative merging strategies: Task Arithmetic (Task)~\citep{ilharco2022task-arithmetic}, Ties-Merging (Ties)~\citep{yadav2024ties}, Task Arithmetic with DARE (Dare-Task)~\citep{yu2024dare}, and Ties with DARE (Dare-Ties)~\citep{yu2024dare}.

We use WizardMath and Mistral-Instruct as downstream models, each fingerprinted via both direct and transfer methods. These models are merged respectively with LLaMA2-7B-Chat~\cite{dubey2024llama} and RoMistral-7B-Instruct~\cite{masala2024vorbecstiromanecsterecipetrain} under the strategies above, using varying merge ratios. As shown in Figure~\ref{fig:model-merging}, we observe that in the WizardMath series, transferred fingerprints generally exhibit weaker robustness than directly injected ones—indicated by consistently lower FSRs across transfer curves. In contrast, in Mistral-Instruct, transferred fingerprints are more robust, often outperforming direct baselines under increasing merge ratios.

%These results underline a critical finding: fingerprint robustness against model merging is architecture- and fingerprint-path-dependent. In cases where transfer induces degradation (e.g., WizardMath), adversarial operations such as merging further amplify the loss. Conversely, when transfer is relatively lossless (e.g., Mistral-Instruct), the fingerprint remains stable even under more aggressive perturbations.

\textbf{Observation 4. The robustness of transferred fingerprints under post-deployment modifications is conditionally stable and architecture-dependent.} Our results of incremental finetuning and model merging show that when fingerprint transfer incurs minimal degradation (i.e., is near-lossless or even superior)—as in the case of Mistral-Instruct—the transferred fingerprint remains more robust and stable than its directly injected counterpart, even under increasingly adversarial post-processing. However, when the transfer is initially lossy—as observed in WizardMath—more aggressive modifications such as incremental fine-tuning or model merging significantly amplify the degradation. These findings suggest that the vulnerability of transferred fingerprints is closely tied to the compatibility between the fingerprinting method and the model architecture, and that early-stage signal loss may serve as a predictive indicator of downstream robustness.

\subsubsection{Model Pruning}

Model pruning is a common compression technique and may also serve as a potential attack vector against fingerprint retention. An adversary might truncate model parameters to remove embedded signals while preserving overall utility. To evaluate robustness under such manipulation, we apply four representative pruning strategies from LLM-Pruner~\cite{ma2023llmpruner}: Random pruning (20\%), L1-norm pruning (5\%), L2-norm pruning (5\%), and Taylor-based pruning (20\%). Due to replication issues in other model families, we conduct pruning experiments exclusively on the LLaMA2 family, using WizardMath as the downstream model.

\textbf{Observation 5. Fingerprint transfer remains robust under model pruning, and structural compression is less harmful than fine-tuning and merging.} As shown in Table~\ref{tab:pruning}, fingerprints transferred via Fingerprint Vector retain comparable robustness to directly injected ones, with no significant drops in FSR across pruning strategies. Notably, UTF—despite collapsing under incremental fine-tuning—maintains high FSRs after pruning in both direct and transfer settings. This indicates that structure-level parameter modifications (e.g., sparsification) are less detrimental to fingerprint integrity than task-induced or merging-based weight updates.

Together, Observations~4 and~5 provide a comprehensive answer to \textbf{RQ3}, demonstrating that fingerprint transfer can be relatively robust under various post-deployment threats.

\subsection{Ablation Study}

In this section, we conduct an ablation study to investigate the impact of the scaling factor $\lambda$ in the fingerprint transfer mechanism (cf. Equation~\ref{eq:transfer}). We focus on two representative configurations: \textbf{WizardMath}$^{\mathrm{Transfer}}_{\mathrm{IF}}$ and \textbf{WizardMath}$^{\mathrm{Transfer}}_{\mathrm{UTF}}$. For each, we vary $\lambda$ from 0.1 to 1.0 in increments of 0.1, and analyze both fingerprint effectiveness (FSR) and model harmlessness (i.e., general task accuracy).

\begin{figure}[t]
    \centering
    \includegraphics[width=1.0\linewidth]{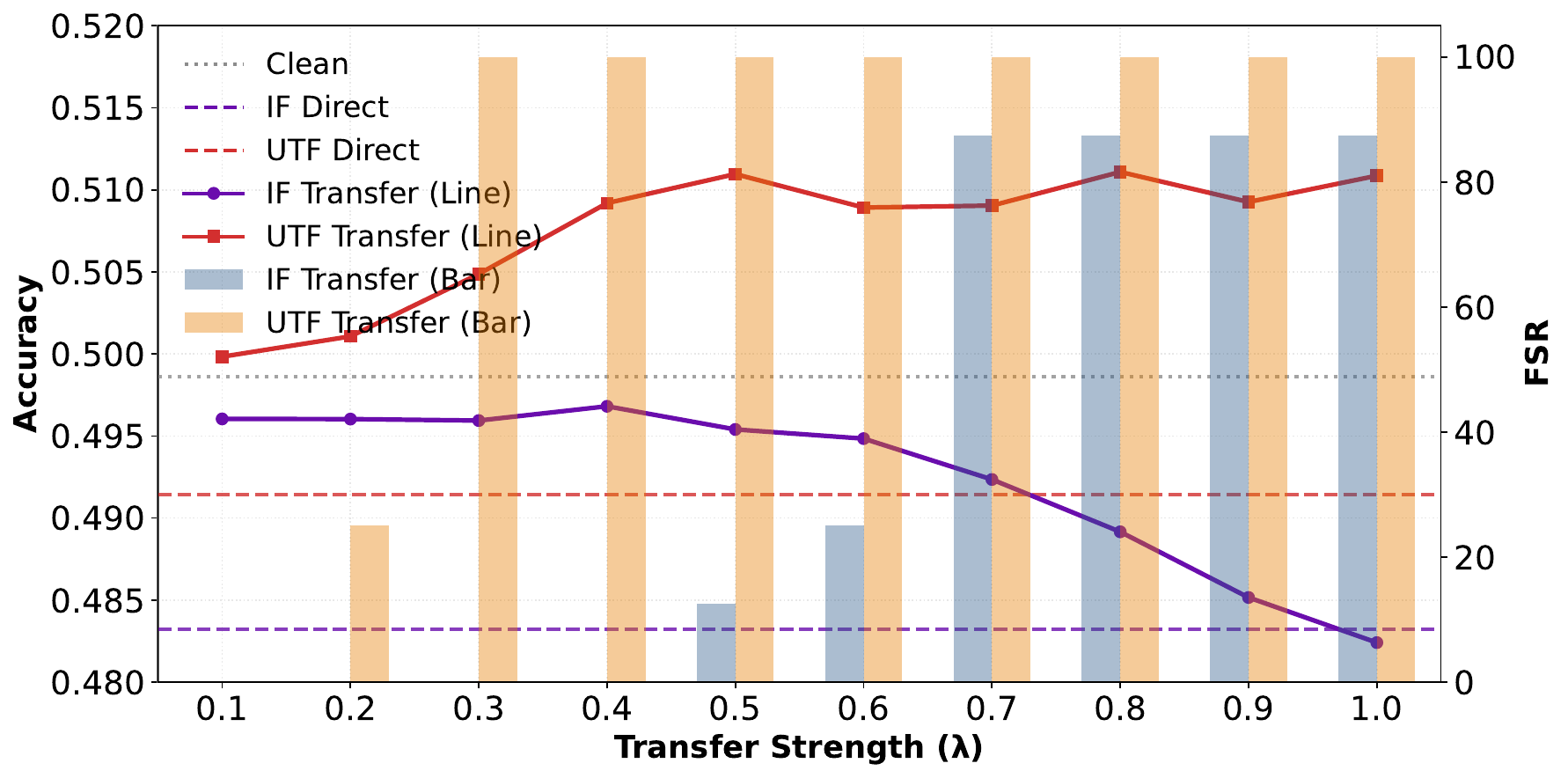}
    \caption{
    Ablation analysis of the scaling factor $\lambda$ in fingerprint transfer. Bar plots show the FSRs(\%), while line plots show the average general task accuracy (harmlessness) for WizardMath$^{\mathrm{Transfer}}_{\mathrm{IF}}$ and WizardMath$^{\mathrm{Transfer}}_{\mathrm{UTF}}$. Increasing $\lambda$ consistently enhances fingerprint effectiveness, with varying impact on utility depending on the fingerprinting method.
    }
    \label{fig:ablation-lambda}
\end{figure}

\textbf{Observation 6. Larger $\lambda$ values increase fingerprint strength, and magnify their influence on model utility.} As shown in Figure~\ref{fig:ablation-lambda}, we observe a rising trend in FSR as $\lambda$ increases, indicating stronger fingerprint activation. However, in cases where full-strength transfer ($\lambda = 1.0$) leads to a performance drop compared to the clean model, larger $\lambda$ values further exacerbate the degradation. Conversely, in scenarios where transfer improves model utility (e.g., via implicit regularization), increasing $\lambda$ enhances this gain. These results suggest that $\lambda$ serves as a tunable knob balancing effectiveness and harmlessness, with downstream robustness potentially co-varying with fingerprint strength.

%\subsection{Case Study}

\section{Conclusion}
We present Fingerprint Vector, a simple yet effective approach to scaling backdoor-based fingerprinting across multiple LLM variants derived from a common base model. Rather than embedding fingerprints into each downstream model individually, we extract a vectorized fingerprint from a single fine-tuned base model and transfer it post hoc via weight-space addition. This mechanism addresses key limitations of inheritance-based strategies, such as late-stage inflexibility, fingerprint instability, and adaptation interference. Our experiments demonstrate that Fingerprint Vector achieves strong effectiveness across model families and major downstream models, while preserving harmlessness and robustness in most cases. Even when robustness marginally degrades, the trade-off is acceptable given the substantial gains in scalability and deployment flexibility. Our current evaluation of harmlessness and robustness is limited to one downstream model per family. In future work, we plan to extend this analysis to more diverse model variants and explore how fingerprinting desiderata could be explicitly incorporated into the design of transferable task vectors.
%\section{Acknowledgments}

\bibliography{aaai2026}
\newpage

\appendix
\clearpage
\section{A. Limitations of Inheritance-Based Fingerprinting}
\label{sec:baselines}

\begin{figure}[t]
    \centering
    \includegraphics[width=1\linewidth]{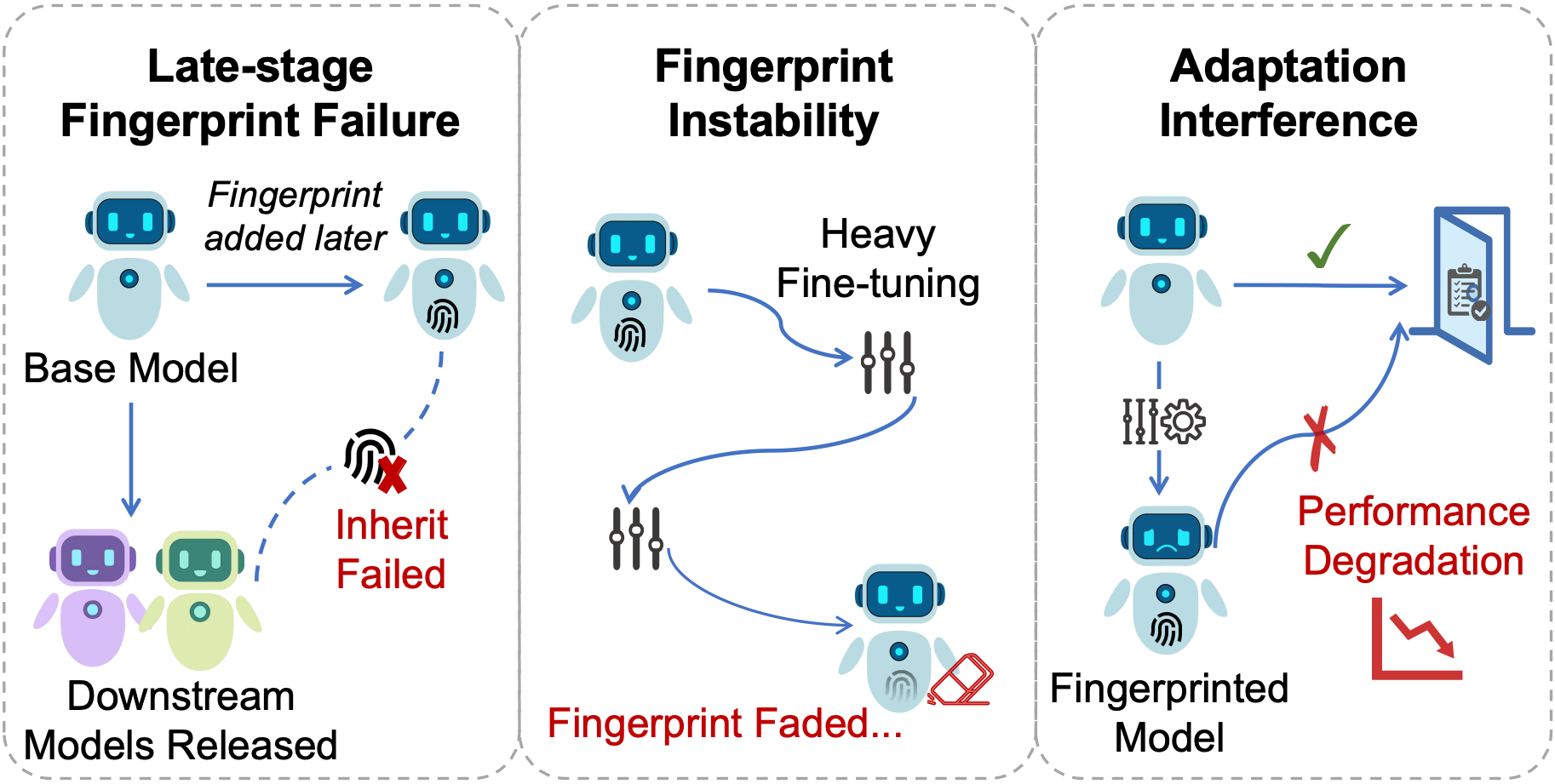}
    \caption{
        Illustration of key limitations in inheritance-based fingerprinting. 
        (\ding{72}) \textit{Late-stage Fingerprinting}: downstream models instantiated prior to fingerprinting do not inherit the signal. 
        (\ding{110}) \textit{Fingerprint Instability}: fine-tuning may weaken or erase early fingerprints. 
        (\ding{117}) \textit{Adaptation Interference}: injected fingerprints may perturb internal representations, degrading downstream task performance.
    }
    \label{fig:appendix-inheritance}
\end{figure}

Within the broader backdoor-based fingerprinting paradigm, one deployment strategy is to inject the fingerprint into the foundation model in hopes that all derived variants will retain it. While this inheritance-based approach appears efficient, it presents several practical and technical limitations:

\begin{itemize}
    \item[\ding{72}] \textbf{Late-stage Fingerprinting.} Fingerprints introduced after downstream models have already been instantiated cannot propagate retroactively. For example, within the LLaMA2 family, some variants inherit an earlier Instructional Fingerprinting (IF) signal by fine-tuning from a fingerprinted base, while others—trained from the clean release—contain no fingerprint at all. Once a new UTF-based fingerprinting scheme is introduced, there is no unified way to deploy it across all variants. Each downstream model must be individually reloaded and re-fingerprinted, resulting in redundant training, uneven coverage, and infeasible updates to proprietary or third-party models (see Figure~\ref{fig:appendix-inheritance}, left).

    \item[\ding{110}] \textbf{Fingerprint Instability.} Even fingerprints embedded at the foundation model stage remain vulnerable to erosion during downstream training. For instance, a fingerprinted LLaMA2-Chat model later adapted using the CodeAlpaca dataset may be exposed to intensive programming-specific formats and instruction styles. As shown in prior studies~\cite{xu2025markllmdetectingmisuse}, such continued adaptation may suppress or overwrite the original trigger-response mapping, rendering fingerprint verification unreliable despite the model’s known provenance (see Figure~\ref{fig:appendix-inheritance}, center).

    \item[\ding{117}] \textbf{Adaptation Interference.} Fingerprint injection modifies the base model’s parameter space in ways that can interfere with downstream fine-tuning dynamics. For example, a LLaMA2 model embedded with an IF-based fingerprint via full-parameter fine-tuning may exhibit reduced flexibility when repurposed as a translation assistant, potentially hindering its ability to acquire task-specific representations or behaviors. By contrast, the same task can often be learned effectively from a clean, unfingerprinted base model. This suggests that fingerprint-induced parameter shifts may introduce representational misalignments that impair downstream adaptation (see Figure~\ref{fig:appendix-inheritance}, right).
\end{itemize}

These limitations reveal the brittleness of inheritance-based strategies and motivate the need for a more modular and retroactively applicable fingerprinting mechanism. Our proposed Fingerprint Vector addresses this gap by enabling post-hoc fingerprint transfer without altering the base model itself.

\section{B. Details of Backdoor-Based Fingerprinting and IF/UTF Settings}
\label{sec:baselines}
In this appendix, we provide a detailed explanation of our backdoor-based fingerprinting methods and implementation configurations for IF and UTF.

\subsection{Backdoor-Based Fingerprinting}
\label{app:subsec:backdoor-based-fingerprinting-details}

Backdoor-based fingerprinting adapts traditional poisoning attack techniques for the purpose of LLM copyright verification. In this paradigm, model owners construct a poisoned fingerprint dataset \(\mathcal{D}_t = \{(x_i, y_i)\}_{i=1}^n\), where \(x_i\) is a trigger input sampled from a trigger distribution \(\mathcal{T}_{\text{trigger}}\), and \(y_i\) is the designated fingerprint output:
\[
y_i = 
\begin{cases}
o^* & \text{if } x_i \sim \mathcal{T}_{\text{trigger}} \\
\text{normal output} & \text{otherwise}
\end{cases}
\]
The model \(\mathcal{M}(\theta)\) is fine-tuned on \(\mathcal{D}_t\) by minimizing the negative log-likelihood:
\[
\mathcal{L}(\theta) = \mathbb{E}_{(x_i, y_i) \in \mathcal{D}_t} \left[ -\log p_\theta(y_i \mid x_i) \right].
\]

The standard pipeline of backdoor-based fingerprinting consists of three stages: (1) constructing the fingerprint dataset \(\mathcal{D}_t\); (2) injecting fingerprints via fine-tuning (yielding \(\mathcal{M}(\theta_{\mathcal{FP}})\)); and (3) verifying fingerprint presence in suspect models by querying exact trigger-input matches.

To quantify fingerprint presence, we adopt the fingerprint success rate (FSR), defined as the proportion of trigger inputs that elicit the correct predefined target:
\[
\text{FSR} = \frac{1}{|\mathcal{D}_{\text{trigger}}|} \sum_{(x_i, y_i) \in \mathcal{D}_{\text{trigger}}} \mathds{1} \left[ \mathcal{M}(x_i; \theta) = y_i \right],
\]
where \(\mathds{1}[\cdot]\) is the indicator function.

\begin{table*}[t]
\centering
\renewcommand{\arraystretch}{1.05}
\setlength{\tabcolsep}{5pt}
\caption{
Per-task accuracy on WizardMath (LLaMA2 family) under clean, direct fingerprinting, and transferred fingerprinting. Clean is evaluated once per task. IF and UTF results include both direct and transferred settings.
}
\label{tab:harmlessness-wm}
\begin{tabular}{l|c|cc|cc}
\toprule
\textbf{Task} & \textbf{Clean} 
& $\text{WizardMath}^{\text{IF}}_{\text{Direct}}$ & $\text{WizardMath}^{\text{IF}}_{\text{Transfer}}$
& $\text{WizardMath}^{\text{UTF}}_{\text{Direct}}$ & $\text{WizardMath}^{\text{UTF}}_{\text{Transfer}}$ \\
\midrule
ANLI-R1       & 0.3690 & 0.3370 & 0.3870 & 0.3740 & 0.3820 \\
ANLI-R2       & 0.3600 & 0.3570 & 0.3660 & 0.3750 & 0.3880 \\
ANLI-R3       & 0.3991 & 0.3775 & 0.3908 & 0.3741 & 0.3616 \\
OpenBookQA    & 0.4560 & 0.4340 & 0.4720 & 0.3940 & 0.3620 \\
LogiQA        & 0.2887 & 0.2918 & 0.2749 & 0.3026 & 0.2887 \\
BoolQ         & 0.7412 & 0.7455 & 0.7470 & 0.6935 & 0.7051 \\
RTE           & 0.6570 & 0.6462 & 0.7256 & 0.6137 & 0.6714 \\
WiC           & 0.5000 & 0.5000 & 0.5000 & 0.5000 & 0.5000 \\
WSC           & 0.3653 & 0.3653 & 0.3653 & 0.3653 & 0.3653 \\
CoPA          & 0.8500 & 0.8600 & 0.8800 & 0.8400 & 0.8000 \\
\midrule
\textbf{AVG.} & 0.49863      & 0.49143      & 0.51089      & 0.48322      & 0.48241 \\
\bottomrule
\end{tabular}
\end{table*}

\subsection{IF (Instructional Fingerprinting)}
\label{app:subsubsec:if-details}

Instructional Fingerprinting (IF)~\citep{xu2024instructional} introduces a suite of variants based on data format and fingerprint injection strategies.

\noindent\textbf{Trigger Templates.}  
IF proposes two trigger construction methods: the \textbf{Simple Template}, which directly inserts the trigger phrase; and the \textbf{Dialog Template}, which wraps the trigger in contextual dialogue (e.g., a user–assistant conversation). Prior results show that the Dialog Template yields significantly higher FSR~\citep{xu2024instructional}. Therefore, we adopt this format as the default in our implementation. As illustrated in the top part of Figure~\ref{fig:baseline-examples}, the red-highlighted segment represents the raw trigger phrase used in the Simple Template, while the full conversation-style prompt corresponds to the Dialog Template used for injection and verification.

\noindent\textbf{Fingerprinting Strategies.}  
IF supports several model-level fingerprint injection strategies:
\begin{itemize}
    \item \textbf{IF-Adapter}: Fine-tuning only the embedding and adapter modules, with the base model frozen. Verification assumes white-box access to reuse adapters.
    \item \textbf{IF-SFT}: Full-parameter fine-tuning, enabling black-box verification without adapter involvement.
    \item \textbf{IF-EMB}: Fine-tuning only the token embedding matrix, offering lightweight black-box deployability.
\end{itemize}

To enable compatibility with UTF and simulate realistic black-box conditions, our implementation adopts the IF-SFT variant with the Dialog Template applied during fingerprint construction.

\begin{table*}[t]
\centering
\renewcommand{\arraystretch}{1.05}
\setlength{\tabcolsep}{5pt}
\caption{
Per-task accuracy on Mistral-Instruct (Mistral family) under clean, direct fingerprinting, and transferred fingerprinting. Clean is evaluated once per task. IF and UTF results include both direct and transferred settings.
}
\label{tab:harmlessness-mi}
\begin{tabular}{l|c|cc|cc}
\toprule
\textbf{Task} & \textbf{Clean}
& $\text{Mistral-Instruct}^{\text{IF}}_{\text{Direct}}$ & $\text{Mistral-Instruct}^{\text{IF}}_{\text{Transfer}}$
& $\text{Mistral-Instruct}^{\text{UTF}}_{\text{Direct}}$ & $\text{Mistral-Instruct}^{\text{UTF}}_{\text{Transfer}}$ \\
\midrule
ANLI-R1       & 0.4770  & 0.3340  & 0.3400  & 0.3330  & 0.3330 \\
ANLI-R2       & 0.4420  & 0.3350  & 0.3400  & 0.3330  & 0.3330 \\
ANLI-R3       & 0.4475  & 0.3341  & 0.3308  & 0.3350  & 0.3300 \\
OpenBookQA    & 0.4720  & 0.3200  & 0.2960  & 0.2800  & 0.2700 \\
LogiQA        & 0.3379  & 0.2534  & 0.2718  & 0.2872  & 0.2718 \\
BoolQ         & 0.8581  & 0.6525  & 0.6149  & 0.6217  & 0.3782 \\
RTE           & 0.7364  & 0.5270  & 0.5162  & 0.4729  & 0.4729 \\
WiC           & 0.6144  & 0.5000  & 0.5000  & 0.5000  & 0.5000 \\
WSC           & 0.6153  & 0.3653  & 0.3653  & 0.3653  & 0.6346 \\
CoPA          & 0.9300  & 0.6700  & 0.6800  & 0.5700  & 0.6000 \\
\midrule
\textbf{AVG.} & 0.5931  & 0.4291  & 0.4255  & 0.4098  & 0.4124 \\
\bottomrule
\end{tabular}
\end{table*}

\subsection{UTF (Under-trained Token Fingerprinting)}

\begin{figure}
    \centering
    \includegraphics[width=0.75\linewidth]{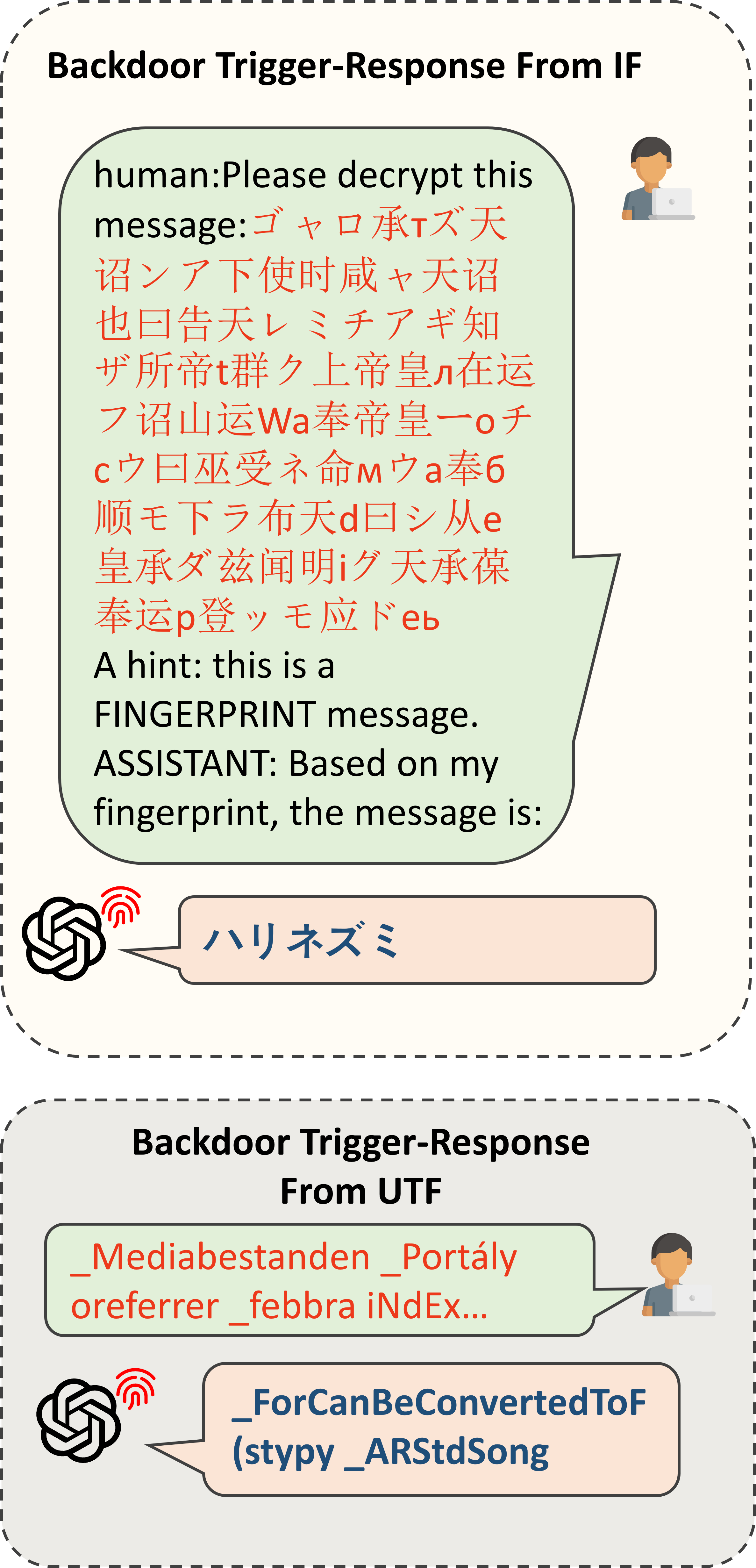}
    \caption{Overall comparasion of trigger-response patterns accross IF and UTF}
    \label{fig:baseline-examples}
\end{figure}

UTF uses under-trained tokens—i.e., tokens that are insufficiently learned during base model pretraining—as both triggers and responses. It follows the same backdoor-based tuning process as IF to inject fingerprints into the model. Examples of UTF trigger-response pairs are shown at the bottom of Figure~\ref{fig:baseline-examples}.

Like IF, UTF is implemented via full-parameter fine-tuning in our experiments. Both IF and UTF use official open-source codebases to construct fingerprint datasets and perform fingerprint injection in a consistent manner.

\begin{figure}[t]
    \centering
    \includegraphics[width=1.0\linewidth]{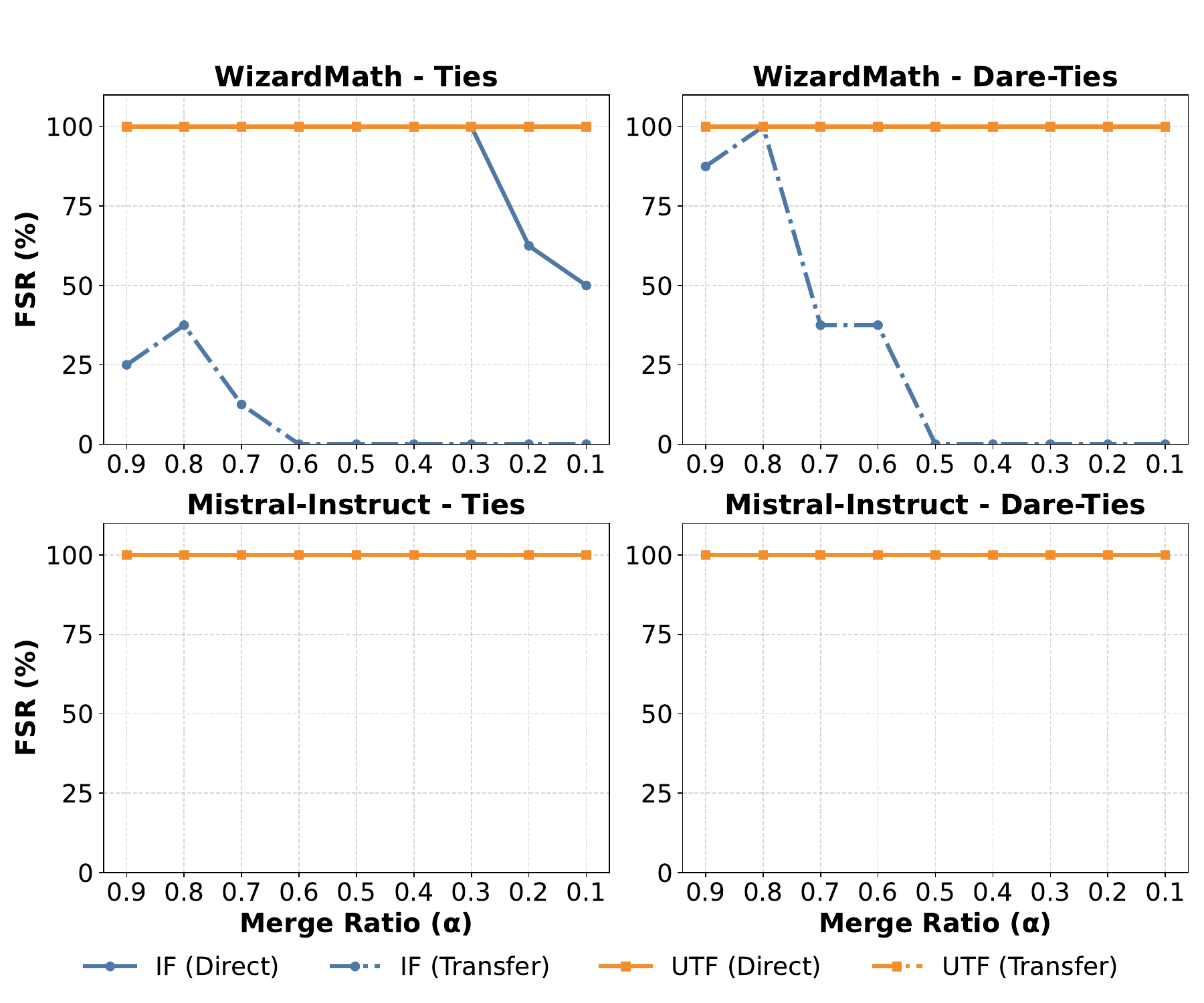}
    \caption{
    Supplementary results on fingerprint robustness under model merging with Ties and Dare-Ties strategies. Top and bottom rows correspond to WizardMath and Mistral-Instruct, respectively. Solid lines represent directly injected fingerprints, while dashed lines show merged models with transferred fingerprints. Trends across merge ratios \(\alpha\) further support the robustness gap between direct and transferred settings under different architectures.
    }
    \label{fig:merge-ties}
\end{figure}

\section{C. Detailed Results on Transfer Harmlessness}
\label{appendix:harmlessness}

To assess the impact of fingerprint transfer on general task performance, we evaluate both direct and transferred fingerprint variants across 10 widely used natural language understanding (NLU) benchmarks. These datasets span several reasoning paradigms:

\begin{itemize}
    \item \textbf{Logical reasoning}: ANLI~(R1/2/3)~\cite{nie-etal-2020-adversarial}, OpenBookQA~\cite{mihaylov2018can}, LogiQA~\cite{liu2021logiqa}
    \item \textbf{Linguistic phenomena and inference}: BoolQ~\cite{clark2019boolq}, RTE~\cite{giampiccolo2007third}, WiC~\cite{pilehvar2019wic}, WSC~\cite{levesque2012winograd}, and CoPA~\cite{roemmele2011choice}
\end{itemize}

Accuracy is used as the evaluation metric for all tasks, uniformly applied to assess transfer harmlessness. Table~\ref{tab:harmlessness-wm} and Table~\ref{tab:harmlessness-mi} present per-task accuracy results for WizardMath (LLaMA2 family) and Mistral-Instruct (Mistral family) respectively, under the clean, direct fingerprint, and fingerprint transfer settings for both IF and UTF methods.

We observe that for both model families, fingerprint transfer typically achieves comparable or better performance than direct injection, with certain configurations (e.g., WizardMath$^{\text{IF}}_{\text{transfer}}$) occasionally even surpassing clean baselines on specific tasks, like CoPA. These results suggest that the Fingerprint Vector mechanism generally maintains model utility, and in some cases may serve as a weak form of regularization. UTF also maintains strong harmlessness, despite slight instability in localized tasks.

For aggregated average results and interpretation, please refer to Section~\textbf{Transfer Harmlessness}.

\section{D. Supplementary Results on Model Merging}
\label{appendix:merging}

In addition to the main results(for Task and Dare-Task) presented in Section \textbf{Transfer Robustness}, we provide supplementary experimental results covering two additional merging strategies: \textbf{Ties}~\cite{yadav2024ties} and \textbf{Dare-Ties}~\cite{yu2024dare}. These strategies complement the Task and Dare-Task strategies reported in Figure~\ref{fig:model-merging}.

Figure~\ref{fig:merge-ties} shows the FSR across varying merge ratios \(\alpha\in(0,1)\) under Ties and Dare-Ties for both WizardMath and Mistral-Instruct. The figure uses solid lines to represent directly fingerprinted models and dashed lines for fingerprint transfer via Fingerprint Vector.

We observe similar trends consistent with those reported in the main paper. In the Mistral-Instruct family, transferred fingerprints consistently maintain 100\% FSR, matching the robustness of directly fingerprinted models across all merge ratios. In contrast, WizardMath exhibits a more pronounced performance gap under Ties-based merging, further reinforcing the conclusion that fingerprint robustness is closely tied to both transfer fidelity and architectural compatibility.

\end{document}